\documentclass{sig-alternate} 
\pdfoutput=1
\usepackage{mathptmx} % This is Times font

\newcommand{\ignore}[1]{}
\usepackage{fancyhdr}
\usepackage[normalem]{ulem}
\usepackage[hyphens]{url}
\usepackage{fixltx2e}
\usepackage[T1]{fontenc}
\usepackage{graphicx}
\usepackage{color}
\usepackage{cite}
\usepackage{tabularx}
\usepackage[inline]{enumitem}
\usepackage{setspace}
\usepackage{epstopdf}
\usepackage{setspace}
\usepackage{color,soul}
\usepackage{parskip}
\usepackage{listings}
\usepackage{caption,subcaption}
\usepackage{layouts}

\usepackage[bookmarks=true,breaklinks=true,letterpaper=true,colorlinks,linkcolor=black,citecolor=blue,urlcolor=black]{hyperref}

\usepackage{xspace}

\newcommand{\hide}[1]{}
\newcommand{\fig}[1]{Figure~\ref{#1}\xspace}
\newcommand{\tbl}[1]{Table~\ref{#1}\xspace}
\newcommand{\sect}[1]{Section~\ref{#1}\xspace}

\newcommand{\ee}[0]{$^{1}$}
\newcommand{\cs}[0]{$^{2}$}
\newcommand{\eecs}[0]{$^{1,2}$}

\newcommand{\tiddelta}[0]{$\Delta TID$\xspace}
\newcommand{\authemail}[1]{\large{\texttt{#1}}}

\fancypagestyle{firstpage}{
	\fancyhf{}
	
	\fancyhead[C]{} 
	\fancyfoot[C]{\thepage}
}  

\title{Inter-thread Communication in Multithreaded, Reconfigurable Coarse-grain Arrays} 
\author{   {Dani Voitsechov\ee ~{and} Yoav Etsion\eecs}\\[0.5ex]
	{\sffamily Electrical Engineering\ee \quad\quad\quad Computer Science\cs}\\
	{\sffamily Technion - Israel Institute of Technology}\\
	\authemail{\{dani,yetsion\}@tce.technion.ac.il}
}
\begin{document}
	\maketitle
	\thispagestyle{firstpage}
	\pagestyle{plain}
	
	\begin{abstract}
		Traditional von Neumann GPGPUs only allow threads to communicate through memory on a group-to-group basis. In this model, a group of producer threads writes intermediate values to memory, which are read by a group of consumer threads after a barrier synchronization. To alleviate the memory bandwidth imposed by this method of communication, GPGPUs provide a small scratchpad memory that prevents intermediate values from overloading DRAM bandwidth.
		
		In this paper we introduce direct inter-thread communications for massively multithreaded CGRAs, where intermediate values are communicated directly through the compute fabric on a point-to-point basis. This method avoids the need to write values to memory, eliminates the need for a dedicated scratchpad, and avoids workgroup-global barriers.
		The paper introduces the programming model (CUDA) and execution model extensions, as well as the hardware primitives that facilitate the communication.
		Our simulations of Rodinia benchmarks running on the new system show that direct inter-thread communication provides an average speedup of 4.5$\times$\ (13.5$\times$\ max) and reduces system power by an average of 7$\times$\ (33$\times$\ max), when compared to an equivalent Nvidia GPGPU.
	
	\end{abstract}
	
	%%%%%%%%%%%%%%%%%%%%%%%%%%
	
	%%%%%%%%%%%%%%%%%%%%%%%%%%%%%%%%%%%%%%%%%%%%%%%%%%%%%%%%%%%%%%%%%%%%%%%%
\section{Introduction}
\label{sec:intro}
%%%%%%%%%%%%%%%%%%%%%%%%%%%%%%%%%%%%%%%%%%%%%%%%%%%%%%%%%%%%%%%%%%%%%%%%

Conventional von Neumann GPGPUs employ the data-parallel single-instruction multiple threads (SIMT) model. But pure data parallelism can only go so far, and the majority of data parallel workloads require some form of thread collaboration through inter-thread communication.
Common GPGPU programming models such as CUDA and OpenCL control the massive parallelism available in the workload by grouping threads into cooperative thread arrays (CTAs; or workgroups). Threads in a CTA share a coherent memory that is used for inter-thread communication.

This model has two major limitations.
The first limitation is that communication is mediated by a shared memory region. As a result, the shared memory region, typically implemented using a hardware scratchpad, must support high communication bandwidth and is therefore energy costly.
The second limitation is the synchronization model. Since the order of scheduling of the threads within a CTA is unknown, a synchronization barrier must be invoked before consumer threads can read the values written to the shared memory by their respective producer threads.

Seeking an alternative to the von Neumann GPGPU model, both the research community and industry began exploring dataflow-based systolic and coarse-grained reconfigurable architectures (CGRA)~\cite{googleTPU,wavecomputingWP,govindaraju11,chen2017eyeriss,nowatzki2016pushing}.
As part of this push, Voitsechov and Etsion introduced the massively multithreaded CGRA (MT-CGRA) architecture~\cite{voitsechov14,voitsechov2015control}, which maps the compute graph of CUDA kernels to a CGRA and uses the dynamic dataflow execution model to run multiple CUDA threads. The MT-CGRA architecture leverages the direct connectivity between functional units, provided by the CGRA fabric, to eliminate multiple von Neumann bottlenecks including the register file and instruction control. This model thus lifts restrictions imposed by register file bandwidth and can utilize all functional units in the grid concurrently. MT-CGRA has been shown to dramatically outperform von Neumann GPGPUs while consuming substantially less power.
Still, the original MT-CGRA model employed shared memory and synchronization barriers for inter-thread communication, and incurred their power and performance overheads.

In this paper we present dMT-CGRA, an extension to MT-CGRA that supports direct inter-thread communication through the CGRA fabric. By extending the programming model, the execution model, and the underlying hardware, the new architecture forgoes the shared memory/scratchpad and global synchronization operations.

The dMT-CGRA architecture relies on the following components:
We extend the CUDA programming model with two primitives that enable programmers to express direct inter-thread dependencies. The primitives let programmers state that thread $N$ requires a value generated by thread $N-k$, for any arbitrary thread index $N$ and a scalar $k$.

The direct dependencies expressed by the programmer are mapped by the compiler to temporal links in the kernel's dataflow graph. The temporal links express dependencies between concurrently-executing instances of the dataflow graph, each representing a different thread.

Finally, we introduce two new functional units to the CGRA that redirect dataflow tokens between graph instances (threads) such that the dataflow firing rule is preserved.

The remainder of this paper is organized as follows. \sect{sec:motivation} describes the motivation for direct inter-thread communication on an MT-CGRA and explains the rationale for the proposed design. \sect{sec:models} then presents the dMT-CGRA execution model and the proposed programming model extensions, and \sect{sec:arch} presents the dMT-CGRA architecture. We present our evaluation in \sect{sec:results} and discuss related work in \sect{sec:related}. Finally, we conclude with \sect{sec:conclusions}.

	%%%%%%%%%%%%%%%%%%%%%%%%%%%%%%%%%%%%%%%%%%%%%%%%%%%%%%%%%%%%%%%%%%%%%
\section{Inter-thread Communication in a multithreaded CGRA}
\label{sec:motivation}
%%%%%%%%%%%%%%%%%%%%%%%%%%%%%%%%%%%%%%%%%%%%%%%%%%%%%%%%%%%%%%%%%%%%%

Modern massively multithreaded processors, namely GPGPUs, employ many von-Neumann processing units to deliver massive concurrency, and use shared memory (a scratchpad) as the primary mean for inter-thread communication. This design imposes two major limitations:
\begin{enumerate}
\item
  The frequency of inter-thread communication barrages shared memory with intermediate results, resulting in high bandwidth requirements from the dedicated scratchpad and dramatically increases its power consumption.
\item
  The asynchronous nature of memory decouples communication from synchronization and forces programmers to use explicit synchronization primitives such as barriers, which impede concurrency by forcing threads to wait until all others have reached the synchronization point.
\end{enumerate}

The dataflow computing model, on the other hand, offers more flexible inter-thread communication primitives. The dataflow model couples direct communication of intermediate values between functional units with the {\it dataflow firing rule} to synchronize computations. We argue that the \emph{massively multithreaded CGRA} (MT-CGRA)~\cite{voitsechov14,voitsechov2015control} design, which employs the dataflow model, can be extended to support inter-thread in the proposed \emph{direct MT-CGRA} (dMT-CGRA) architecture.
Specifically, inter-thread communication on the \emph{dMT-CGRA} architecture is  implemented as follows. Whenever an instruction in thread $A$ sends a data token to an instruction in thread $B$, the latter will not execute (i.e., fire) until the data token from thread $A$ has arrived. This simple use of the dataflow firing rule ensures that thread $B$ will wait for thread $A$.
The dataflow model therefore addresses the two limitation of the von Neumann model:
\begin{enumerate}
\item
  The CGRA fabric, with its internal buffers, enables most communicated values to propagate directly to their destination, thereby avoiding costly communication with the shared memory scratchpad. Only a small fraction of tokens that cannot be buffered in the fabric are spilled to memory.
\item
  By coupling communication and synchronization using message-passing extensions to SIMT  programming models, dMT-CGRA implicitly synchronizes point-to-point data delivery without costly barriers.
\end{enumerate}

The remainder of this section argues for the coupling of communication and synchronization, and discusses why typical programs can be satisfied by the internal CGRA buffering.

%%%%%%%%%%%%%%%%%%%%%%%%%%%%%%%%%%%%%%%%%%%%%%%%%%%%%%%%%%%%%%%%%%%%%
\subsection{Dataflow and message passing}
\label{sec:value_passing}
%%%%%%%%%%%%%%%%%%%%%%%%%%%%%%%%%%%%%%%%%%%%%%%%%%%%%%%%%%%%%%%%%%%%%

%%%%%%%%%%%%%%%%%%%%%%%%%%%%%%%%%%%%%%%%%%%%%%%%%%
\lstset {
	language=C++,
	basicstyle=\ttfamily\scriptsize,
	breaklines=false,
	commentstyle=\color{red}\scriptsize\it,
        morekeywords={For,each},
        keywordstyle=\bf,        
}
%%%%%%%%%%%%%%%%%%%%%%%%%%%%%%%%%%%%%%%%%%%%%%%%%%

%%%%%%%%%%%%%%%%%%%%%%%%%%%%%%%%%%%%%%%%%%%%%%%%%%
\begin{figure}[t!]
  \centering

\begin{subfigure}[b]{\columnwidth}
\centering
\begin{lstlisting}[frame=tb]
thread_code(thread_t tid) {

  // common: not next to margin
  if(!is_margin(tid - 1) &&
     !is_margin(tid + 1) ) {
    result[tid] = globalImage[tid-1] * kernel[0]
                + globalImage[tid]   * kernel[1]
                + globalImage[tid+1] * kernel[2];
  }
  // corner: next to left margin
  else if(is_margin(tid - 1)) {
    result[tid] = globalImage[tid-1] * kernel[0]
                + globalImage[tid]   * kernel[1];
  }
  // corner: next to right margin
  else if(is_margin(tid - 1)) {
    result[tid] = globalImage[tid]   * kernel[1];
                + globalImage[tid+1] * kernel[2];
  }
}
\end{lstlisting}
\vspace*{-2ex}
\caption{Spatial convolution using only global memory}
\vspace*{1ex}
\label{fig:convolution_global}
\end{subfigure}
%
%
%
% new line of figures  
\begin{subfigure}[b]{\columnwidth}
\centering
\begin{lstlisting}[frame=tb]
thread_code() {
  // map the thread to 1D space (CUDA-style)
  tid = threadIdx.x;

  // load image into shared memory
  sharedImage[tid] = globalImage[tid];

  // pad the margins with zeros
  if (is_margin(tid))
    pad_margin(sharedImage, tid);

  // block until all threads finish the load phase
  barrier(): // e.g. CUDA syncthreads 

  // execute the convolution; (kernel
  // (preloaded in shared memory)
  result[tid] =
      sharedImage[tid-1] * kernel[0]
    + sharedImage[tid  ] * kernel[1]
    + sharedImage[tid+1] * kernel[2];
}
\end{lstlisting}
\vspace*{-2ex}
\caption{Spatial convolution on a GPGPU using shared memory}
\vspace*{1ex}
\label{fig:convolution_GPU}
\end{subfigure}
%
%\vspace*{2ex}
%
\begin{subfigure}[b]{\columnwidth}
\centering
\begin{lstlisting}[frame=tb]  
thread_code() {
  // map the thread to 1D space (CUDA-style)
  tid = threadIdx.x;
  
  // load one elemnt from global memory
  mem_elem = globalImage[tid];

  // tag the value of the variable to be sent,
  // in case the variable gets rewritten.
  tagValue<mem_elem>();
  
  // wait for tokens from threads tid+1 and tid-1
  lt_elem = fromThreadOrConst<mem_elem,/*tid*/-1,0>();
  rt_elem = fromThreadOrConst<mem_elem,/*tid*/+1,0>();

  // execute the convolution
  result[tid] = lt_elem  * kernel[0]
              + mem_elem * kernel[1]
              + rt_elem  * kernel[2];
}              
\end{lstlisting}
\vspace*{-2ex}
\caption{Spatial convolution on a MT-CGRA using thread cooperation}
\vspace*{1ex}
\label{fig:convolution_SGMF}
\end{subfigure}
\caption{Implementation of a \emph{separable convolution}~\cite{podlozhnyuk2007image}) using various inter-thread data sharing models. For brevity, we focuses on 1D-convolutions, which are the main and iterative component in the algorithm.}
\label{fig:convolution}
\end{figure}
%%%%%%%%%%%%%%%%%%%%%%%%%%%%%%%%%%%%%%%%%%%%%%%%%%

We demonstrate our dMT-CGRA message-passing extensions using a \emph{separable convolution} example~\cite{podlozhnyuk2007image} that is included with the NVIDIA software development kit (SDK)~\cite{nvidiasdk}.
This convolution applies a kernel to an image by applying  1D-convolutions on each image dimension. For brevity, we focus our discussion on a single 1D-convolution with a kernel of size 3. The example is depicted using pseudo-code in \fig{fig:convolution}. 

Separable convolution can be implemented using global memory, shared memory, or a message-passing programing model.
The trivial parallel implementation, presented in \fig{fig:convolution_global}, uses global memory. If the entire kernel falls within the image margins, the matrix elements should be simply multiplied with the corresponding elements of the convolution kernel. If either thread IDs (TID) \emph{TID - 1} or \emph{TID + 1} are outside the margins, their matching element should be zero. Although this naive implementation is very easy to code, it results in multiple memory assesses to the image, which are translated to high power consumption and low performance.

GPGPUs use shared memory to overcome this issue, as shown in \fig{fig:convolution_GPU}. The code first loads each element of the matrix once and stores it in the shared memory (sharedImage array in the code), and the image in shared memory is padded with zeros. A barrier synchronization must then be used to ensure that all threads finished loading their values to shared memory. Only after the barrier can the actual convolution be computed. Nevertheless, although the computation phase now accesses shared memory, the lack of direct inter-thread communication forces redundant accesses, as each image and kernel element is loaded by multiple threads.

A dataflow architecture, on the other hand, can seamlessly incorporate a message passing framework for inter-thread communication. \fig{fig:convolution_SGMF} demonstrates how separable convolution can be implemented in dMT-CGRA.
The fundamental message passing primitive in dMT-CGRA is that threads are allowed to request the values of other threads' variables. Given that the underlying single-instruction multiple-threads (SIMT) model, threads are homogeneous and execute the same code (with diverging control paths).

As \fig{fig:convolution_SGMF} show, each thread first loads one matrix element to a register (as opposed to the shared memory write in \fig{fig:convolution_GPU}). Once the element is loaded, the thread goes on to wait for values read from other threads.
The programmer must tag the version of the named variable (in case the variable is rewritten) that should be available to remote threads using the \emph{tagValue} call. 
Thread(s) can then read the remote value using the \emph{fromThreadOrConst()} call (see \sect{sec:api} for the full API), which takes three arguments: the name of the remote variable, the thread ID from which the value should be read, and a default value in case the thread ID is invalid (e.g., a negative thread ID). Similar to CUDA and OpenCL, thread IDs are mapped to multi-dimensional coordinates (e.g., \emph{threadIdx} in CUDA~\cite{cuda2015programming}) and Thread IDs are encoded as constant deltas between the source thread ID and the executing thread's ID.

Communication calls are therefore translated by the compiler to edges in the code's dataflow graph representing dependencies between instances of the graph (i.e., threads). This process folds the data transfers into the underlying dataflow firing rule (to facilitate compile-time translation, the arguments are passed as C++ template parameters).

This implicit embedding of the communication into the dataflow graph gives the model its strengths.
Primarily, the dataflow graph enables the dMT-CGRA processor to forward values between threads directly, eliminating the need for shared-memory mediation.
In addition, the embedding allows threads to move to the computation phase once their respective values are ready, independently of other threads. Since no barriers are required, the implicit dataflow synchronization does not impede parallelism.

%%%%%%%%%%%%%%%%%%%%%%%%%%%%%%%%%%%%%%%%%%%%%%%%%%%%%%%%%%%%%%%%%%%%%
\subsection{Forwarding memory values between threads}
\label{sec:mem_value_passing}
%%%%%%%%%%%%%%%%%%%%%%%%%%%%%%%%%%%%%%%%%%%%%%%%%%%%%%%%%%%%%%%%%%%%%

%%%%%%%%%%%%%%%%%%%%%%%%%%%%%%%%%%%%%%%%%%%%%%%%%%
\begin{figure}[t!]
%  \vspace*{-3ex}
\centering
\begin{subfigure}[b]{\columnwidth}
\centering
\begin{lstlisting}[frame=tb]
// IDs in a thread block are mapped to
// 2D space (e.g., CUDA)
thread_code() {
  // map the thread to 2D space (CUDA-style)
  tx = threadIdx.x;
  ty = threadIdx.y;

  // load A and B into shared memory
  sharedA[tx][ty] = A[tx][ty];
  sharedB[tx][ty] = B[tx][ty];

  // block until all threads finish the load phase
  barrier(): // e.g. CUDA syncthreads 

  // compute an element in sharedC (dot product)
  sharedC[tx][ty] = 0;
  for(i=0; i<K; i++)
    sharedC[tx][ty] += sharedA[tx][i]*sharedB[i][ty];
  }

  // write back dot product result to global memory
  C[tx][ty] = sharedC[tx][ty]
}
\end{lstlisting}
\vspace*{-2ex}
\caption{Matrix multiplication on a GPGPU using shared memory.}
\label{fig:matMul_GPU}
\vspace*{1ex}
\end{subfigure}

\begin{subfigure}[b]{\columnwidth}
\centering
\begin{lstlisting}[frame=tb]
thread_code() {
  // mapping the thread to 2D space (CUDA-style)
  tx = threadIdx.x;
  ty = threadIdx.y;
  
  //compute memory access predicates
  En_A = (tx == 0);
  En_B = (ty == 0);

  // compute the dot product. the loop is statically
  // unrolled to compute the indices a compile-time
  C[ty][tx] = 0;
#pragma unroll
  for(i=0; i<K; i++) {
    a = fromThreadOrMem<{0, -1}>(A[tx][i], En_A);
    b = fromThreadOrMem<{1,  0}>(B[i][ty], En_B);

    C[ty][tx] += a*b;
  }
}
\end{lstlisting}
\vspace*{-2ex}
\caption{Dense matrix multiplication on the dMT-CGRA architecture using direct inter-thread communication.}
\label{fig:matMul_SGMF}
\vspace*{1ex}
\end{subfigure}

\vfill
\caption{Multiplications of dense matrices $C = A \times B$ using shared memory on GPGPU and direct inter-thread communication on an MT-CGRA. Matrix dimensions are $(N\times M) = (N\times K) \times (K\times M)$.}
\label{fig:matMul}
\vspace*{-2ex}
\end{figure}

\begin{figure}[t]
\centering
\includegraphics[width=\columnwidth]{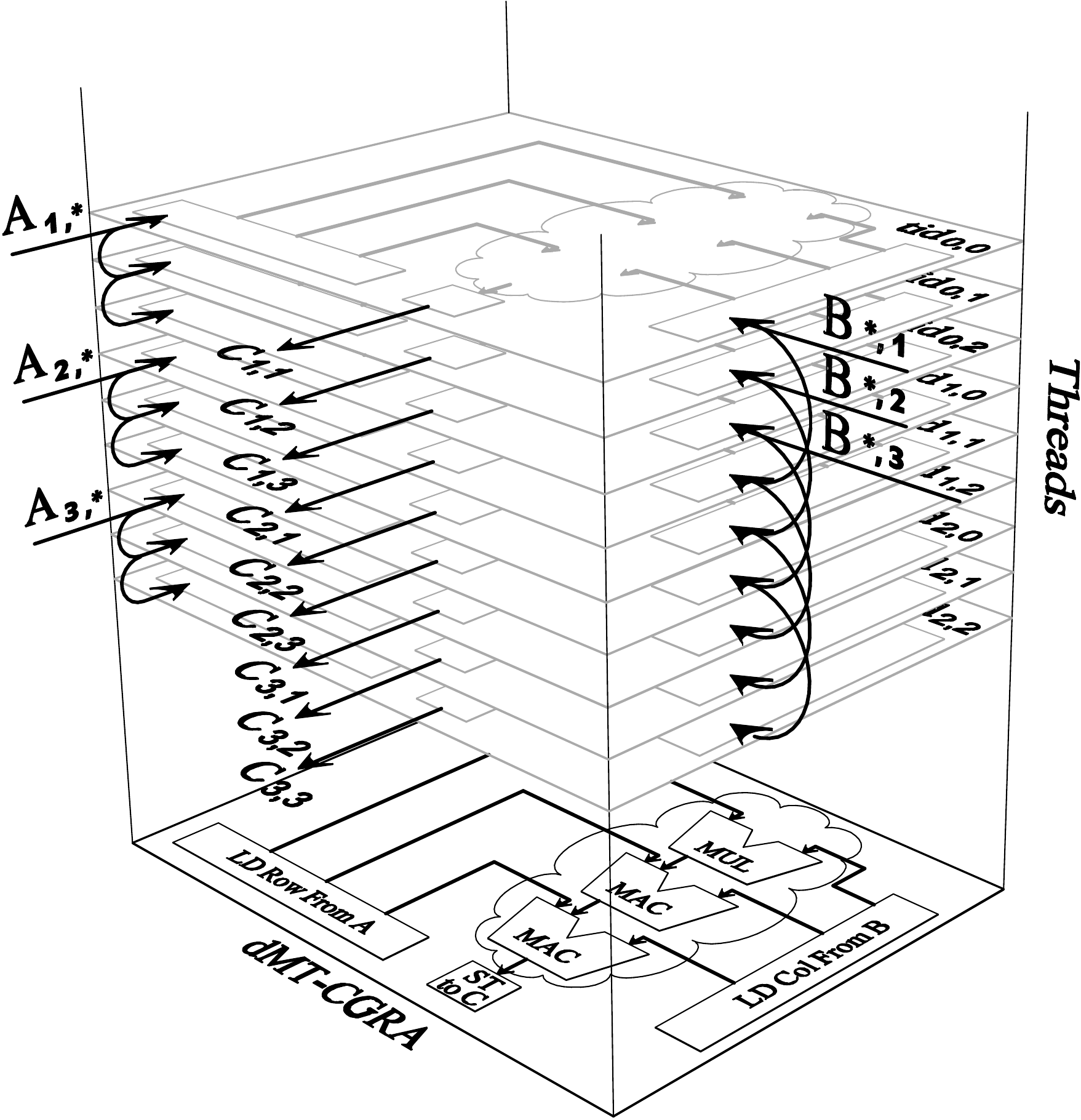}
\vspace*{0.1ex}
\caption{The flow of data in dMT-CGRA for a 3x3 marix multiplication. The physical CGRA is configured with the dataflow graph (bottom layer), and each functional unit in the CGRA multiplexes operations from different instances (i.e., threads) of the same graph.}
\label{fig:matMul_diag}
%\end{subfigure}
\end{figure}
%

%%%%%%%%%%%%%%%%%%%%%%%%%%%%%%%%%%%%%%%%%%%%%%%%%%

Often times multiple concurrent threads load the same multiple address, thereby  stressing the memory system with redundant loads. The synergy between a CGRA compute fabric and direct inter-thread communication enables dMT-CGRA to
forward values loaded from memory through the CGRA fabric, and thus eliminate most of these redundant loads.

\fig{fig:matMul} presents this property using matrix multiplication as an example. The figure depicts the implementation of a dense matrix multiplication $C = A \times B$ on a GPGPU and on dMT-CGRA. In both implementations each thread computes one element of the result matrix $C$.

The classic GPGPU implementation, shown in \fig{fig:matMul_GPU}, demonstrates how multiple threads stress memory. The implementation concurrently copies the data from global memory to shared memory, executes a synchronization barrier (which impedes parallelism), and then each thread computes one element in the result matrix $C$. Consequently, each element in the source matrices $A$ and $B$, whose dimensions are $N\times K$ and $K \times M$, respectively, is accessed by all threads that compute a target element in $C$ whose coordinates  correspond to either its row or column. As a result, each element is loaded by $N \times M$ threads.

We propose to eliminate the redundant memory accesses by introducing a new memory-or-thread communication primitive. The new primitive uses a compile-time predicate that determines whether to load the value from memory or to forward the loaded value from another thread. The dMT-CGRA toolchain maps the operation to special units in the CGRA (described in \sect{sec:arch}) and, using the predicate, configures the CGRA to route the correct value.

\fig{fig:matMul_SGMF} depicts an implementation of a dense matrix multiplication using the proposed primitive. Each thread in the example computes one element in the destination matrix $C$, and the programming model maps each thread to a spatial coordinate (similar to CUDA/OpenCL). Rather than a regular memory access, the code uses the \emph{fromThreadOrMem} primitive, which takes two arguments: a predicate, which determines where to get the value from, a memory address, from which the required value should be loaded, and one parameter a two-dimensional coordinate, which indicate the thread from which the data may be obtained (the coordinates are encoded as the multi-dimensional difference between the source thread and the executing thread's coordinates).

Finally, \fig{fig:matMul_diag} illustrates the flow of data between threads for a $3 \times 3$ matrix multiplication. While the figure shows a copy of the dataflow graph for each thread (we remind the reader that the underlying dMT-CGRA is configured with a single dataflow graph and executes multiple threads by moving their tokens through the graph out-of-order, using dynamic dataflow token-matching).
As each thread computes one element in target matrix $C$, threads that compute the first column load the elements of matrix A from memory, and the threads that compute the first row load the elements of matrix $B$. As the figure shows, threads that load values from memory forward them to other threads.
For example, thread $(0,2)$ loads the bottom row of matrix A and forwards its values to thread $(1,2)$, which in turn sends them to thread $(2,2)$.
Since thread $(2,2)$.

The combination of a multithreaded CGRA and direct inter-thread communication thus greatly alleviates the load on the memory system, which plagues massively parallel processors. The following sections elaborate on the design of the  programming model, the dMT-CGRA execution model, and the underlying architecture.

	%%%%%%%%%%%%%%%%%%%%%%%%%%%%%%%%%%%%%%%%%%%%%%%%%%%%%%%%%%%%%%%%%%%%%
\section{Execution and programming model}
\label{sec:models}
%%%%%%%%%%%%%%%%%%%%%%%%%%%%%%%%%%%%%%%%%%%%%%%%%%%%%%%%%%%%%%%%%%%%%
This section describes the dMT-CGRA execution model and the programming model  extensions that support direct data movement between threads.

%%%%%%%%%%%%%%%%%%%%%%%%%%%%%%%%%%%%%%%%%%%%%%%%%%
\paragraph*{\bf The MT-CGRA execution model}\quad
The MT-CGRA execution model combines the \emph{static} and \emph{dynamic} dataflow models to execute single-instruction multiple-threads (SIMT) programs with better performance and power characteristics than von Neumann GPGPUs~\cite{voitsechov14}.
The model converts SIMT kernels into dataflow graphs and maps them to the CGRA fabric, where each functional unit multiplexes its operation on tokens from different instances of a dataflow graph (i.e., threads).

An MT-CGRA core comprises a host of interconnected functional units (e.g., arithmetic logical units, floating point units, load/store units). Its architecture is described in Section~\ref{sec:arch}.
The interconnect is configured using the program's dataflow graph to statically move tokens between the functional units. Execution of instructions from each graph instance (thread) thus follows the \emph{static dataflow model}. In addition, each functional unit in the CGRA employs \emph{dynamic, tagged-token dataflow}~\cite{arvind90,patt85} to dynamically schedule different threads' instructions in order to prevent memory stalled threads from blocking other threads, thereby maximizing the utilization of the functional units.

Prior to executing a kernel, the functional units and interconnect are  configured to execute a dataflow graph that consists of one or more replicas of the kernel's dataflow graph. Replicating the kernel's dataflow graph enables the architecture to better utilize the MT-CGRF grid.
The configuration process itself is lightweight and has negligible impact on system performance. Once configured, threads are streamed through the dataflow core by injecting their thread identifiers and CUDA/OpenCL coordinates (e.g.,  threadIdx in CUDA) into the array. When those values are delivered as operands to successor functional units they initiate the thread's computation, following the dataflow firing rule. A new thread can thus be injected into the computational fabric on every cycle.

%%%%%%%%%%%%%%%%%%%%%%%%%%%%%%%%%%%%%%%%%%%%%%%%%%
\paragraph*{\bf Inter-thread communication on an MT-CGRA}\quad
As described above, the MT-CGRA execution model is based on \emph{dynamic, tagged-token dataflow}. Each token is coupled with a tag so that functional units can match each thread's input tokens. The multithreaded model uses TIDs as token tags, which allows each functional unit to match each thread's input tokens. When using this model, the crux of inter-thread communication is changing a token's tag to a different TID.

%%%%%%%%%%%%%%%%%%%%%%%%%%%%%%%%%%%%%%%%%%%%%%%%%%
\begin{figure}[t!]
  \centering
\begin{lstlisting}[frame=tb,escapeinside={(*}{*)}]
// return the tagged-token for a given tid
<token, tag> = elevator_node(tid)
{
  // does the source tid falls within the thread block?
  if(in_block_boundaries(tid - (*$\Delta$*))) {
    // valid source tid? wait for the token.
    token = wait_for_token(tid - (*$\Delta$*));
    return <token, tid>;
  }
  else {
    // invalid source tid? push the constant value.
    return <C, tid>;
  }
}
\end{lstlisting}
\caption{The functionality of an elevator node (with a $\Delta$ TID shift and a fallback constant C.}
\label{fig:elevator_abstract}
\end{figure}
%%%%%%%%%%%%%%%%%%%%%%%%%%%%%%%%%%%%%%%%%%%%%%%%%%

We implement the token re-tagging by adding special \emph{elevator} nodes to the CGRA. Like an elevator, which shifts people between floors, the elevator node shifts tokens between TIDs. An elevator node is a single-input, single-output node and is configured with two parameters --- a \tiddelta and a constant $C$. The functionality of the node is described as pseudo code in \fig{fig:elevator_abstract} (and is effectively the implementation of the \emph{fromThreadOrConst} function first described in \fig{fig:convolution_SGMF}). For each downstream thread id $TID$, the node generates a tagged-token consisting of the value obtained from the input token for thread ID $TID-\Delta$. If $TID-\Delta$ is not a valid TID in the thread block, the downstream token consists of a preconfigured constant $C$.

The elevator node thus communicates tokens between threads whose TIDs differ by $\Delta$, which is extracted at compile-time from either the \emph{fromThreadOrConst} or \emph{fromThreadOrMem} family of functions (\sect{sec:api}).
These inter-thread communication functions are mapped by the compiler to elevator nodes in the dataflow graph and to their matching counterparts in the CGRA.

Each elevator node includes a small token buffer. This buffer serves as a single-entry output queue for each target TID. The \tiddelta that a single elevator node can support is thus limited by the token buffer size. To support \tiddelta that are larger than a single node's token buffer, we design the elevator nodes so that they can be cascaded, or chained. Whenever the compiler identifies a \tiddelta that is larger than a single elevator node's token buffer, it maps the inter-thread communication operation to a sequence of cascading elevator nodes. In extreme cases where \tiddelta is too large even for multiple cascaded nodes, dMT-CGRA falls back to spilling the communicated values to the shared memory. Cascading of elevator nodes is further discussed in \sect{sec:arch}.

%%%%%%%%%%%%%%%%%%%%%%%%%%%%%%%%%%%%%%%%%%%%%%%%%%
\begin{figure}[t]
\centering
\includegraphics[width=\linewidth]{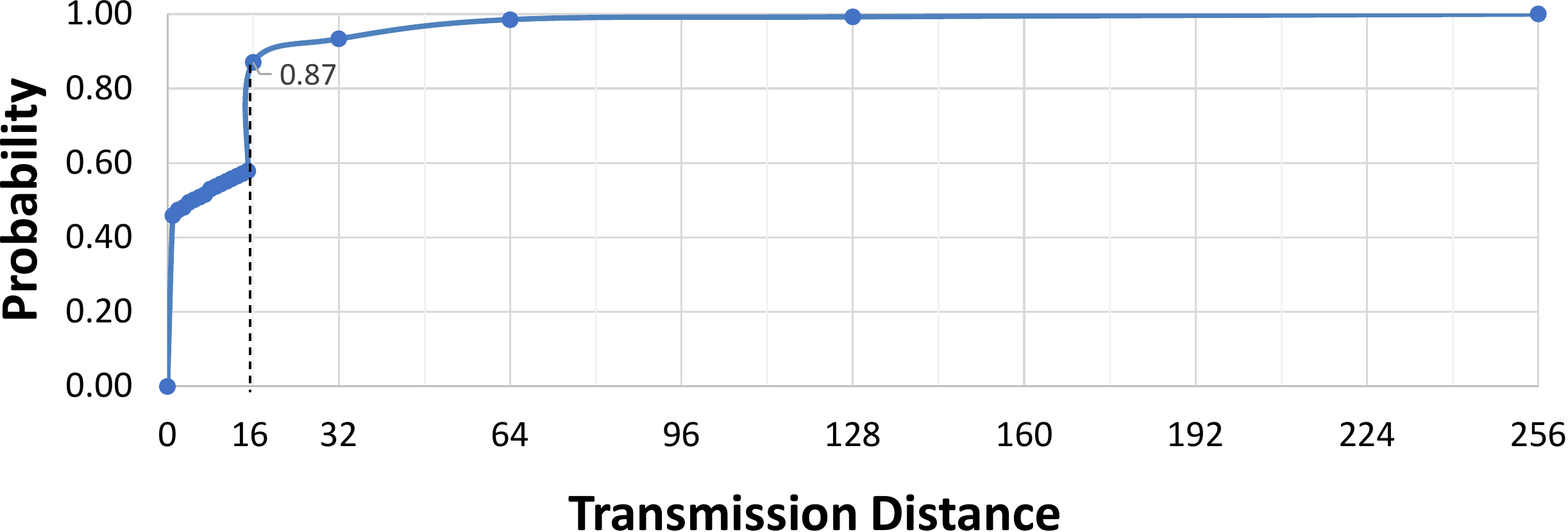}
\vspace{0.5ex}
\caption{cumulative distribution function (CDF) of delta lengths across various benchmarks. We see the 87\% of the code we evaluated uses communicates across \tiddelta of 16, indicating strong communicates locality.}
\label{fig:CDF_deltas}
\end{figure}
%%%%%%%%%%%%%%%%%%%%%%%%%%%%%%%%%%%%%%%%%%%%%%%%%%

Nonetheless, our experimental results show that inter-thread communication patterns exhibit locality across the TID space, and that values are typically communicated between threads with adjacent TID (a Euclidean distance was used for 2D and 3D TID spaces).
Figure~\ref{fig:CDF_deltas} shows the cumulative distribution function (CDF) of the \tiddelta's exhibited by the benchmarks used in this paper (the benchmarks and methodology are described in \sect{sec:method}). The figure shows that the commonly used delta are small and a token buffer of 16 is enough to support 87\% of the benchmark without the need to cascade {\it elevator} nodes.

The second functional unit needed to implement inter-thread communication is the enhanced load/store (eLDST) unit. The eLDST extends a regular LDST unit with a predicated bypass, allowing it to return values coming either from memory or from another thread (through an elevator unit). An eLDST units coupled with an elevator unit (or multiple thereof) thus implement the \emph{fromThreadOrMem} primitive.

%%%%%%%%%%%%%%%%%%%%%%%%%%%%%%%%%%%%%%%%%%%%%%%%%%%%%%%%%%%%%%%%%%%%%
\subsection{Programming model extensions}
\label{sec:api}
%%%%%%%%%%%%%%%%%%%%%%%%%%%%%%%%%%%%%%%%%%%%%%%%%%%%%%%%%%%%%%%%%%%%%

%%%%%%%%%%%%%%%%%%%%%%%%%%%%%%%%%%%%%%%%%%%%%%%%%%
\begin{table*}
  \scriptsize
  \centering
  \begin{tabular}{|l|l|}
    \hline
    Function & Description 						\\
\hline
        {\tt token fromThreadOrConst<variable, TID$\Delta$, constant>()}
        & Read \texttt{variable} from another thread,
          or \texttt{constant} if the thread does not exist.		\\
        {\tt token fromThreadOrConst<variable, TID$\Delta$, constant, win>()}
        & Same as above, but limit the communication to a window of \texttt{win}
        threads.							\\
        \texttt{void tagValue<variable>()}
        & Tag a variable value that will be send to another thread. 	\\
\hline
\hline
        {\tt token fromThreadOrMem<TID$\Delta$>(address, predicate)}
        & Load \texttt{address} if \texttt{predicate} is true,
          or get the value from another thread. 			\\
        {\tt token fromThreadOrMem<TID$\Delta$, win>(address, predicate)}
        & Same as above, but limit the communication to a window of \texttt{win}
        threads.							\\
\hline    
  \end{tabular}
  \vspace*{3pt}
  \caption{API for inter-thread communications. Static/constant values are passed as template parameters (functions that require \tiddelta have versions for 1D, 2D, and 3D TID spaces).}
  \label{tab:api}
\end{table*}
%%%%%%%%%%%%%%%%%%%%%%%%%%%%%%%%%%%%%%%%%%%%%%%%%%

We enable direct inter-thread communication by extending the CUDA/OpenCL API. The API, listed in \tbl{tab:api}, allows threads to communicate with any other thread in a thread block. In this section we describe the three components of the API.

%%%%%%%%%%%%%%%%%%%%%%%%%%%%%%%%%%%%%%%%%%%%%%%%%%%%%%%%%%%%%%%%%%%%%
\subsection{Communicating intermediate values}
\label{sec:rcvFrom_sendTo}
%%%%%%%%%%%%%%%%%%%%%%%%%%%%%%%%%%%%%%%%%%%%%%%%%%%%%%%%%%%%%%%%%%%%%

%%%%%%%%%%%%%%%%%%%%%%%%%%%%%%%%%%%%%%%%%%%%%%%%%%
\begin{figure}[t]
\centering
\begin{minipage}[b]{.58\linewidth}
\begin{subfigure}[b]{\linewidth}
\centering
\includegraphics[width=0.5\linewidth]{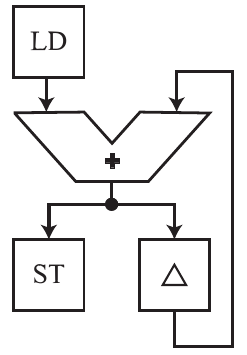}
\caption{The static dMT-CGRA mapping when executing prefix sum.}
\label{fig:prefix_sum_static}
\end{subfigure}
\begin{subfigure}[b]{\linewidth}
\begin{lstlisting}[frame=tb]
thread_code() {
  // mapping the thread to
  // 1D space (CUDA-style)
  tid = threadIdx.x;
	
  //load one value (LD) 
  //from global memory
  mem_val = inArray[tid];

  //add the loaded value to 
  //the sum so far
  sum = 
   fromThreadOrConst<sum,-1,0>() 
   + mem_val;
  tagValue<sum>();
	
  //store partial sum to 
  //global memory 
  prefixSum[tid] = sum;
}
\end{lstlisting}
\caption{Prefix sum implementation using inter thread communication.}
\label{fig:prefix_sum_code}
\end{subfigure}
\end{minipage}
\hfill
\begin{subfigure}[b]{0.35\linewidth}
  \centering
\includegraphics[width=0.8\linewidth,trim={0 0 0 0},clip]{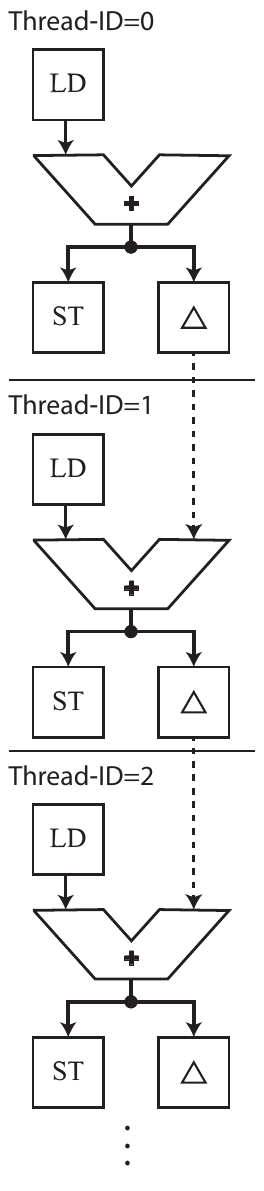}
\caption{The dynamic execution of prefix sum.}
\label{fig:prefix_sum_dynamic}
\end{subfigure}
\vspace{0.5ex}
\caption{Example use of the \emph{tagValue} function.}
\label{fig:tagvalue}
\end{figure}
%%%%%%%%%%%%%%%%%%%%%%%%%%%%%%%%%%%%%%%%%%%%%%%%%%

The \emph{fromThreadOrConst} and \emph{tagValue} functions enables threads to communicate intermediate values in a producer-consumer manner. The function is mapped to one or more elevator nodes, which send a tagged-token downstream once the sender thread's token is received. This behavior blocks the consumer thread until the producer thread sends the token.
The \emph{fromThreadOrConst} function has two variants. The variants share three template parameters: the name of the \emph{variable} to be read from the sending thread, the \emph{\tiddelta} between the communicating threads (which may be multi dimensional), and a \emph{constant} to be used if the sending TID is invalid or outside the \emph{transmission window}.

The \emph{transmission window} is defined as the span of TIDs that share the communication pattern. The second variant of the \emph{fromThreadOrConst} function allows the programmer to bound the window using the \emph{win} template parameter.
More concretely, we define the \emph{transmission window} as follows:
The \emph{fromThreadOrConst} functions encodes a monotonic communication pattern between threads, e.g., thread \emph{TID} produces a value to thread TID+$\Delta$, which produces a value to thread \emph{TID+$2\times\Delta$}, and so forth. The transmission window is defined as the maximum difference between TIDs that participate in the communication pattern. For a window of size \emph{win}, the thread block will be partitioned into consecutive thread groups of size \emph{win}, e.g., threads $[TID_{0} \dots TID_{win-1}]$, $[TID_{win} \dots TID_{2\times win-1}]$, and so on. The communication pattern $TID \rightarrow TID+\Delta$ will be confined to each group, such that (for each $n$) thread $TID_{n \times win-1}$ will not produce a value, and thread $TID_{n \times win}$ will receive the default constant value rather than wait for thread $TID_{n \times win-\Delta}$.

Bounding the transmission window is useful to group threads at the sub-block level. In our benchmarks (\sect{sec:method}), for example, we found grouping useful for computing reduction trees. A bounded transmission window enables mapping distinct groups of communicating threads to separate segments at each level of the tree.

The \emph{tagValue} function is used to tag a specific value (or version) of the  variable passed to \emph{fromThreadOrConst}. The call to \emph{tagValue} may be placed before or after the call to \emph{fromThreadOrConst}, as shown in the \emph{prefix sum} example depicted in \fig{fig:tagvalue} (the example is based on the NVIDIA CUDA SDK~\cite{nvidiasdk}). The prefix sum problem takes an array $a$ of values and, for each element $i$ in the array, sums the array values $a[0]\dots a[i]$.
The code in \fig{fig:prefix_sum_code} uses the \emph{tagValue} to first compute an element's prefix sum, which depends on the value received from the previous thread, and only then send the result to the subsequent thread. \fig{fig:prefix_sum_static} illustrates the resulting per-thread dataflow graph, and \fig{fig:prefix_sum_dynamic} illustrates the inter-thread communication pattern across multiple threads (i.e, graph instances). The resulting pattern demonstrates how decoupling the \emph{tagValue} call from the \emph{fromThreadOrConst} call allows the compiler to schedule the store instruction in parallel with the inter-thread communication, thereby exposing more instruction-level parallelism (ILP). 

%%%%%%%%%%%%%%%%%%%%%%%%%%%%%%%%%%%%%%%%%%%%%%%%%%%%%%%%%%%%%%%%%%%%%
\subsection{Forwarding memory values}
\label{sec:fromThreadOrMem}
%%%%%%%%%%%%%%%%%%%%%%%%%%%%%%%%%%%%%%%%%%%%%%%%%%%%%%%%%%%%%%%%%%%%%

The \emph{fromThreadOrMem} function allows threads that load the same memory address to share a single load operation. The function takes \emph{\tiddelta} as a template parameter, and an \emph{address} and \emph{predicate} as run time evaluated parameters (the function also has a variant that allows the programmer to bound the transmission window). Using the predicate, the function can dynamically determine which of the threads will issue the actual load instruction, and which threads will \emph{piggyback} on the single load and get the resulting value forwarded to them. An typical use of the \emph{fromThreadOrMem} function is shown in the matrix multiplication example in \fig{fig:matMul_SGMF}. In this example, the function allows for only a single thread to load each row and each column in the matrices, and for the remaining threads to receive the loaded value from that thread. In this case, the memory forwarding functionality reduces the number of memory accesses from $N \times K \times M$ to $N \times M$.

	%%%%%%%%%%%%%%%%%%%%%%%%%%%%%%%%%%%%%%%%%%%%%%%%%%
\lstset {
	language=C++,
	basicstyle=\ttfamily\scriptsize,
	breaklines=false,
	commentstyle=\color{red}\scriptsize\it,
	morekeywords={For,each},
	keywordstyle=\bf,    
	escapeinside={<@}{@>},    
}
%%%%%%%%%%%%%%%%%%%%%%%%%%%%%%%%%%%%%%%%%%%%%%%%%%

%%%%%%%%%%%%%%%%%%%%%%%%%%%%%%%%%%%%%%%%%%%%%%%%%%%%%%%%%%%%%%%%%%%%%
\section{The \lowercase{d}MT-CGRA Architecture}
\label{sec:arch}
%%%%%%%%%%%%%%%%%%%%%%%%%%%%%%%%%%%%%%%%%%%%%%%%%%%%%%%%%%%%%%%%%%%%%

%%%%%%%%%%%%%%%%%%%%%%%%%%%%%%%%%%%%%%%%%%%%%%%%%%
\begin{figure}[t!]
	\centering
	\begin{subfigure}[b]{\linewidth}
		\includegraphics[width=\linewidth,trim={1.5cm 5.1cm 1.5cm 0},clip]{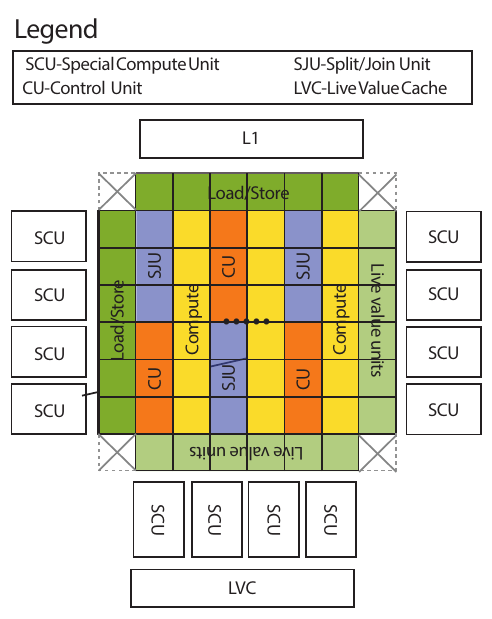}
		\caption{An SGMF MT-CGRA core}
		\label{fig:full_architecture}
	\end{subfigure}

	\begin{subfigure}[b]{\linewidth}
		\includegraphics[width=\linewidth,trim={1.5cm 0 1.5cm 6.2cm},clip]{full_architecture_typical-unit}
		\caption{A typical MT-CGRA unit}
		\label{fig:typical_node}
	\end{subfigure}
	\caption{MT-CGRA core overview }
	\label{fig:MT-CGRA_architecture}
\end{figure}
%%%%%%%%%%%%%%%%%%%%%%%%%%%%%%%%%%%%%%%%%%%%%%%%%%

This section describes the dMT-CGRA architecture, focusing on the extensions to the baseline MT-CGRA~\cite{voitsechov14} needed to facilitate inter-thread communication. \fig{fig:MT-CGRA_architecture} illustrates the high-level structure of the MT-CGRA architecture.

The MT-CGRA core itself, presented in \fig{fig:full_architecture}, is a grid of functional units interconnected by a statically routed network on chip (NoC). The core configuration, the mapping of instructions to functional units, and NoC routing are determined at compile-time and written to the MT-CGRA when the kernel is loaded. During execution tokens are passed between the various functional units according to the static mapping of the NoC. The grid is composed of  heterogeneous functional units, and different instructions are mapped to different unit types in the following manner: Memory operations are mapped to the load/store units, computational operations are mapped to the floating point units and ALUs (compute units), control operations such as select, bitwise operations and comparisons are mapped to control units (CU), and split and join operations (used to preserve the original intra-thread memory order) are mapped to Split/Join units (SJU).

During the execution of parallel tasks on an MT-CGRA core, many different flows representing different threads reside in the grid simultaneously. Thus, the information is passed as tagged tokens composed from the data itself and the associated TID, which serves as tag. The tag is used by the grid's nodes to determine which operands belong to which threads.

\fig{fig:typical_node} illustrates the shared structure of the different units. While the funcnionality of the units differ, they all include tagged-token matching logic to support thread interleaving through dynamic dataflow. Specifically, tagged-tokens arrive from the NoC and are inserted into the token buffer. Once a all operands for a specific TIDs are available, they are passed to the unit's logic (e.g., access memory in LDST units, compute in ALU/FPU). When the unit's logic complete its operation, the result is passed as a tagged token back to the grid through the unit's crossbar switch.      

In this paper we introduce two new units to the grid --- the \emph{elevator node} and the \emph{enhanced load/store unit} (eLDST). While existing units may manipulate the the token itself, they to not modify the tag in order to preserve the association between tokens an threads. The two new units facilitate inter-thread communication by modifying the tags of existing tokens.

\fig{fig:elevator_node} and \fig{fig:eLDST_unit} depict the elevator node and eLDST units, respectively. Nevertheless, we introduce the new units to the grid by converting the existing control units to elevator nodes and LDST units to eLDST units. 
The conversion only includes adding combinatorial logic to the existing units, since all units in the grid already have an internal opcode register and token buffers.
The $\mu$architectural overhead of the conversion thus consists of only the combinational logic shown in \fig{fig:elevator_node} and \fig{fig:eLDST_unit}.
Consequently, the conversion of existing units incurs negligible area and power overhead.

%%%%%%%%%%%%%%%%%%%%%%%%%%%%%%%%%%%%%%%%%%%%%%%%%%
\subsection{Elevator node}
%%%%%%%%%%%%%%%%%%%%%%%%%%%%%%%%%%%%%%%%%%%%%%%%%%

%%%%%%%%%%%%%%%%%%%%%%%%%%%%%%%%%%%%%%%%%%%%%%%%%%
\begin{figure}[t!]
	\centering
	\begin{subfigure}[b]{\linewidth}
		\includegraphics[width=\linewidth]{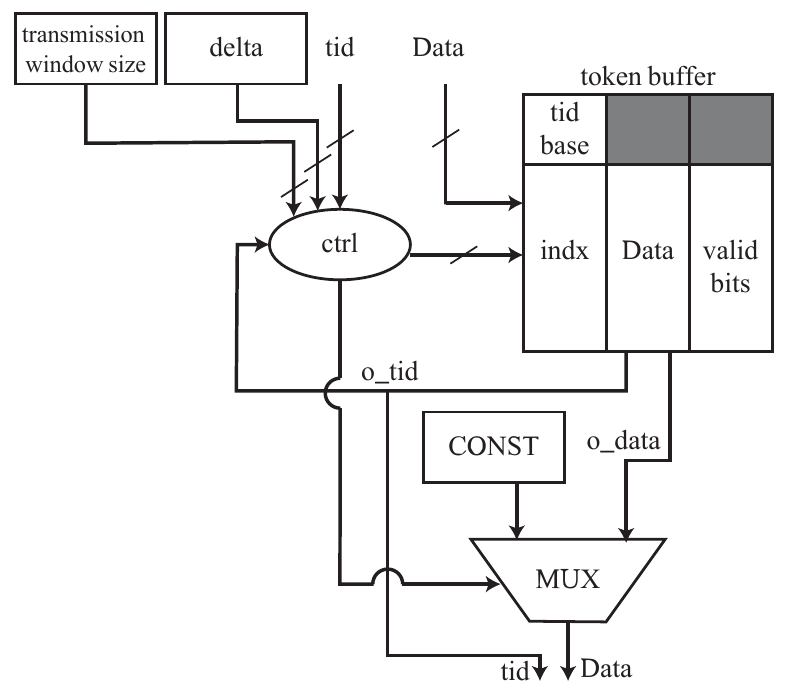}
		\caption{An elevator node stores the in-flight tokens in the unit's token buffer. A controller manipulates TIDs and controls the value of the output tokens.}
		\label{fig:elevator_blocks}
	\end{subfigure}

\begin{subfigure}[b]{\columnwidth}
	\centering
	\begin{lstlisting}[frame=tb]
//read a token from the node's input and 
//manipulate the tag
if(tid%BATCH < DELTA)
{ //for threads acting only as producers
	token_buffer.validate(tid)
	token_buffer.add(tid+DELTA,data)
} 
else if(tid%BATCH + DELTA < BATCH)
{ //for threads acting as produsers and consumers
	token_buffer.add(tid+DELTA,data)
}


//if the token buffer holds a ready thread
//pop it's token and test whether it should 
//return a value or a const 
if(!token_buffer.isEmpty()){
	(o_data,o_tid) = token_buffer.pop()
	if(o_tid%BATCH >= DELTA) 
	//the thread has a producer
		out<=(o_tid,o_data)
	else 
	//generate a const value for threads 
	//without producers
		out<=(o_tid,CONST)
}
	\end{lstlisting}
	%%%%%%%%%%%%%%%%%%%%%%%%%%%%%%%%%%%%%%%%%% with positive and negative deltas%%%%%%%%%
%%%%%%%%%%%%%%%%%%%%%%%%%%%%%%%%%%%%%%%%%%%%%%%%%%%%%%%%%%%%%%%%%%%%%%%%%%%%%%%%%%%%%%%%%%%%%%
	\vspace*{-2ex}
	\caption{Pseudo-code for the \emph{elevator} node controller when treating \tiddelta greater than zero.}
	\vspace*{1ex}
	\label{fig:elevator_controller}
\end{subfigure}
		\vspace*{-2ex}
		\caption{The elevator node and its controller's pseudo-code.}
		\label{fig:elevator_node}
		
\end{figure}
%%%%%%%%%%%%%%%%%%%%%%%%%%%%%%%%%%%%%%%%%%%%%%%%%%

%%%%%%%%%%%%%%%%%%%%%%%%%%%%%%%%%%%%%%%%%%%%%%%%%%
%%%%%%%%%%%%%%%%%%%%%%%%%%%%%%%%%%%%%%%%%%%%%%%%%%

The elevator node implements the \emph{fromThreadOrConst} function, which  communicates intermediate value between threads, and is depicted in \fig{fig:elevator_node}. When mapping \emph{fromThreadOrConst} call, the node is configured with the call's \tiddelta and default constant value. An \emph{elevator} node receives tokens tagged with a $TID$ and changes the tag to $TID+\Delta$ according to its preconfigured $\Delta$. It then sends the resulting tagged-token downstream.

\fig{fig:elevator_controller} shows the pseudo-code for the node's functionality. 
In the most common case, the node receives an input token from one thread and sends the retagged token to another thread. In this case, threads serve as both data producers, sending a token to a consumer thread, and as consumers, waiting for a token from another producer thread.
Alternatively, a thread $TID$ may not serve as a producer if its target thread's ID $TID+\Delta$ is invalid or outside the current transmission window.
Correspondingly, when the sending thread's ID (e.g., $TID-\Delta$) is outside the transmission window, the elevator node injects the preconfigured constant to the tagged-token sent downstream.

\fig{fig:elevator_blocks} depicts the structure of an elevator unit.
For threads acting that both produce and consume tokens, the controller passes the input token to its receiver by modifying the tag from $TID$ to $TID+\Delta$, and pushing the resulting tagged-token to the $TID+\Delta$ entry in the token buffer. In addition, the original input $TID$ should be acknowledged by marking the thread as ready in the token buffer.
Alternatively, if thread $TID$ simply needs to receive the pre-defined constant value as a token, the controller pushes a tagged-token comprising the constant and $TID$ to the token buffer. In this case, setting the acknowledged bit does not require an extra write port to the token buffer but only the ability to turn two bits at once.

%%%%%%%%%%%%%%%%%%%%%%%%%%%%%%%%%%%%%%%%%%%%%%%%%%
\subsection{Enhanced load/store unit (eLDST)}
%%%%%%%%%%%%%%%%%%%%%%%%%%%%%%%%%%%%%%%%%%%%%%%%%%

%%%%%%%%%%%%%%%%%%%%%%%%%%%%%%%%%%%%%%%%%%%%%%%%%%
\begin{figure}[t!]
	\includegraphics[width=\linewidth]{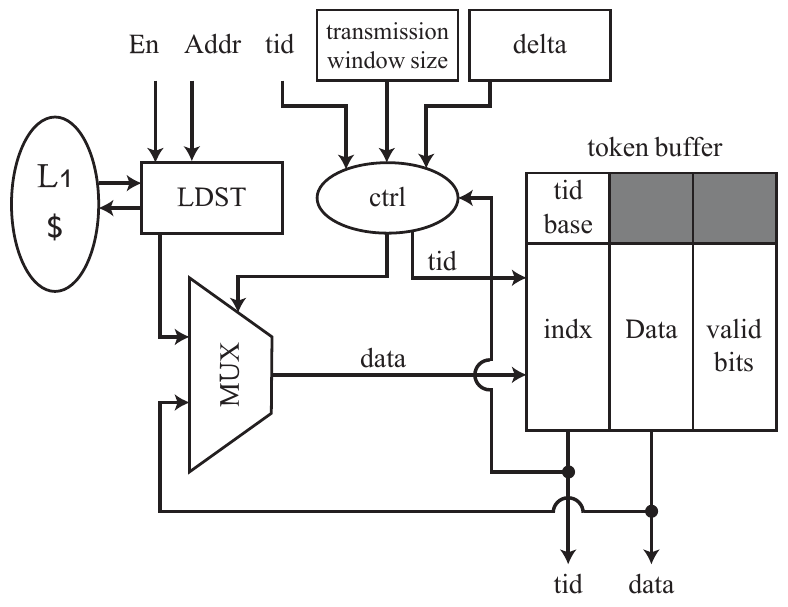}
	\caption{An eLDST node, comprising a LDST unit with additional adder to manipulate the TID, and a comparator to test if the result is outside the margins. The \emph{En} input (predicate) determines whether a new value should be introduced. The output is looped back in to create new tokens with higher TIDs with the same data loaded by a previous thread.}
	\label{fig:eLDST_unit}
\end{figure}
%%%%%%%%%%%%%%%%%%%%%%%%%%%%%%%%%%%%%%%%%%%%%%%%%%

The eLDST unit is used to implement the \emph{fromThreadOrMem} function, which enables threads to reuse memory values loaded another thread without issuing redundant memory accesses. Figure~\ref{fig:eLDST_unit} presents the eLDST unit, which is a LDST unit enhanced with control logic that determines whether the token should be brought in from memory or from another thread's slot in the token buffer.
The eLDST units operates as follows: if the \emph{Enable} (En) input is set, the receiving thread will access the memory system and load the data. Otherwise, if the En is not set the thread's TID will either be added to the token buffer where it will wait for another thread to write the token, or the controller will find the token holding the data fetched from memory waiting in the token buffer. In the latter scenario, the thread may continue its flow through the dataflow graph. When the eLDST produces an output token, the token is duplicated and one copy is internally parsed by the node's logic.
While the original token is passed on downstream in the MT-CGRA, $\Delta$ is added to the TID of the duplicated token. If the resulting TID is equal or smaller than the \emph{transmission window}, the tagged-token will be push to the token buffer. Otherwise, the duplicated token will be discarded since it's consumer is outside the \emph{transmission window}. Using this scheme, each value is loaded once from memory and reused $\frac{window size}{\Delta}$ times, thereby significantly reducing the memory bandwidth.

%%%%%%%%%%%%%%%%%%%%%%%%%%%%%%%%%%%%%%%%%%%%%%%%%%
\subsection{Supporting large transmission distances}
%%%%%%%%%%%%%%%%%%%%%%%%%%%%%%%%%%%%%%%%%%%%%%%%%%

%%%%%%%%%%%%%%%%%%%%%%%%%%%%%%%%%%%%%%%%%%%%%%%%%%
\begin{figure}[t!]
	\centering
	\begin{subfigure}[b]{\linewidth}
		\includegraphics[width=\linewidth,trim={0, 0, 0.8cm, 0}, clip]{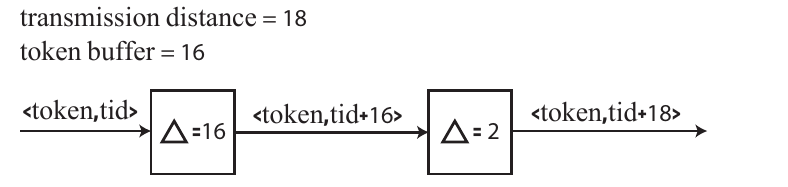}
		\caption{Cascading elevator nodes to manage a \tiddelta that is larger than the token buffer.}
		\label{fig:elevator_chain}
	\end{subfigure}
	
	\begin{subfigure}[b]{\linewidth}
		\includegraphics[width=\linewidth]{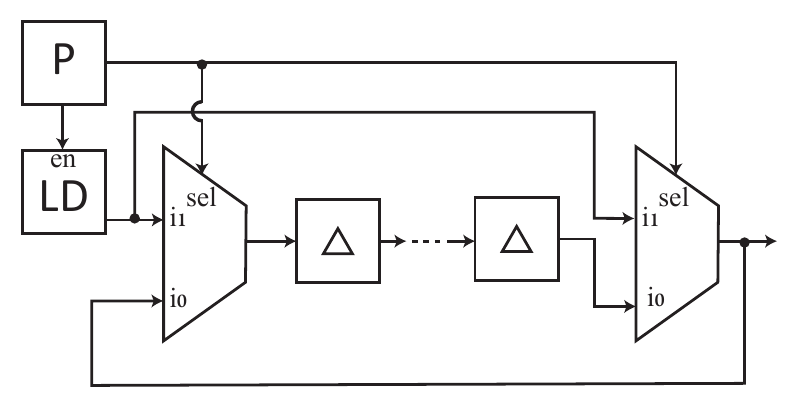}
		\caption{When a \emph{fromThreadOrMem} procedure needs to deal with \emph{$\Delta$} larger than the token buffer size, the function will be mapped to a cascade of predicated \emph{elevator} nodes in a closed cycle.}
		\label{fig:elevator_cycle}
	\end{subfigure}
	\caption{Cascading \emph{elevator} nodes}
	\label{fig:big_deltas}
\end{figure}
%%%%%%%%%%%%%%%%%%%%%%%%%%%%%%%%%%%%%%%%%%%%%%%%%%

%%%%%%%%%%%%%%%%%%%%%%%%%%%%%%%%%%%%%%%%%%%%%%%%%%
%%%%%%%%%%%%%%%%%%%%%%%%%%%%%%%%%%%%%%%%%%%%%%%%%%

The dMT-CGRA architecture uses the token buffers in \emph{elevator} nodes and \emph{eLDST} units to implement inter-thread communication. During compilation, the compiler examines the distance between the sending thread and the receiving thread represented as the \tiddelta passed to the \emph{fromThreadOrConst} or \emph{fromThreadOrMem} functions. If the distance is smaller or equal to the size of the token buffer, the \emph{fromThreadOrConst} or \emph{fromThreadOrMem} calls will be mapped to a single \emph{elevator} node or \emph{eLSDT} unit, respectively. But if \tiddelta is larger than the token buffer, the compiler must cascade multiple nodes to support the large transmission distance.

%%%%%%%%%%%%%%%%%%%%%%%%%%%%%%%%%%%%%%%%%%%%%%%%%%
\paragraph*{\bf Long distances in \emph{fromThreadOrConst} calls}\quad
When a \emph{fromThreadOrConst} function needs to communicate values over a transmission distance that is larger than the size of the token buffer, the compiler cascades multiple \emph{elevator} nodes (effectively chaining their token buffers) in order to support the required communication distance.

\fig{fig:elevator_chain} depicts such a scenario. The required transmission distance shown in the figure is 18, but the token buffer can only hold 16 entries. To deal with the long distance the compiler maps the operation to two cascaded \emph{elevator} nodes. The compiler further configures the \tiddelta of the first node to 16 (the token buffer size) and that of the second one to 2, resulting in the desired cummulative transmission distance of 18.

In the general case of a \emph{transmission window} that is larger than the token buffer size, the number of cascaded units will be
$\left \lceil{\frac{TID\Delta}{Token\ Buffer\ Size}}\right \rceil$. In extreme cases, where the \tiddelta is so large that it requires more elevator nodes that are available in the CGRA, the communicated values will be spilled to the \emph{Live Value Cache}, a compiler managed cache used in the MT-CGRA architecture~\cite{voitsechov2015control}. This approach is similar to the spill fill technique used in GPGPUs.

Nevertheless, spilling values is a rarity since typical transmission distance are small, as shown in \fig{fig:CDF_deltas}. The figure shows the cumulative distribution function (CDF) of the transmission distances in the benchmarks used in this paper. As the CDF shows, the commonly used distances are small and a token buffer of 16 is sufficient to support 87\% of the benchmarks without the need to cascade \emph{elevator} nodes.

%%%%%%%%%%%%%%%%%%%%%%%%%%%%%%%%%%%%%%%%%%%%%%%%%%
\paragraph*{\bf Long distances in \emph{fromThreadOrMem} procedures}\quad
By default, \emph{fromThreadOrMem} calls are mapped to eLDST units. Unlike the \emph{elevator} node that can be cascaded to increase the maximal transmission  distance, the eLDST can not simply be cascaded since it acts as a local buffer for its in-flight memory accesses. For example, in \fig{fig:matMul_diag} the columns of matrix {\bf B} are loaded by the first three threads and transmitted over a distance of three threads ($\Delta TID=3$). In this case, while the third thread loads its data, the eLDST must be able to hold on the first two loaded values in order to transmit them later. As a result, a token buffer of at-least three entries is required. A system with a token buffer smaller than that would require external buffering. 

The additional external buffer is constructed by mapping the operation to a loop of cascaded \emph{elevator} nodes. As depicted in \fig{fig:elevator_cycle}, the loop is enclosed by \emph{control} nodes serving as MUXs. To reuse memory values by distant threads the output of the terminating MUX is connected to the input of the first MUX. In this scenario the compiler will map the load instruction to a predicated load-store unit. The predicate passed to the \emph{fromThreadOrMem} will serve as the selector for the MUXs. When the predicate evaluates to false, the original memory value is looped back through the second MUX back to the \emph{elevator} nodes cascade. A value originating from the TID entering the cascade will be retagged with the target thread ID $TID + \sum_{i}\Delta_i$. The sum of the \emph{elevator} node \tiddelta therefore accounts for the required communication distance.
Nevertheless, as shown in \fig{fig:CDF_deltas}, the typical \tiddelta fits inside the eLDST unit's token buffer.

	%%%%%%%%%%%%%%%%%%%%%%%%%%%%%%%%%%%%%%%%%%%%%%%%%%%%%%%%%%%%%%%%%%%%%
\section{Evaluation}
\label{sec:results}
%%%%%%%%%%%%%%%%%%%%%%%%%%%%%%%%%%%%%%%%%%%%%%%%%%%%%%%%%%%%%%%%%%%%%

%%%%%%%%%%%%%%%%%%%%%%%%%%%%%%%%%%%%%%%%%%%%%%%%%%%%%%%%%%%%%%%%%%%%%
\subsection{Methodology}
\label{sec:method}
%%%%%%%%%%%%%%%%%%%%%%%%%%%%%%%%%%%%%%%%%%%%%%%%%%%%%%%%%%%%%%%%%%%%%

The amount of logic that is found in a dMT-CGRA core is approximately the same amount that is found in an Nvidia SM and in an SGMF MT-CGRA core. In an Nvidia SM, that logic assembles 32 CUDA cores, while in the dMT-CGRA core the Nvidia SM is broken down into smaller coarse grained blocks. The breakdown to a smaller granularity exposes instruction-level parallelism (ILP) while the use of multithreading preserves the thread level parallelism (TLP).

%%%%%%%%%%%%%%%%%%%%%%%%%%%%%%%%%%%%%%%%%%%%%%%%%%
\begin{table}[t]
	\vspace*{-1.5ex}
	\footnotesize
	\centering
	\begin{tabular}{| l | l |}
		\hline
		Parameter & Value \\
		\hline \hline
		dMT-CGRA Core & 140 interconnected\\ 
		& compute/LDST/control units\\
		\hline
		Arithmetic units  & 32 ALUs \\
		Floating point units  & 32 FPUs \\
		& 12 Special Compute units \\
		Load/Store units & 32 LDST Units \\
		Control units      & 16 Split/Join units \\
		& 16 Control/Elevator units\\
		\hline
		Frequency [GHz] & core 1.4, Interconnect 1.4  \\
		&   L2 0.7, DRAM 0.924 \\
		\hline
		L1 & 64KB, 32 banks, 128B/line, 4-way\\
		L2 & 786KB, 6 banks, 128B/line, 16-way\\
		GDDR5 DRAM & 16 banks, 6 channels \\
		\hline
	\end{tabular}
	\caption{dMT-CGRA system configuration.\label{tab:cache_conf}}
\end{table}
%%%%%%%%%%%%%%%%%%%%%%%%%%%%%%%%%%%%%%%%%%%%%%%%%%

%\vspace{-1.5ex}
\paragraph*{\bf Simulation framework}\quad
We used the GPGPU-Sim simulator~\cite{gpgpusim} and GPUWattch~\cite{leng13} power model (which uses performance monitors to estimate the total execution energy) to evaluate the performance and power of the dMT-CGRA, the MT-CGRA and the GPUs architecture.
These tools model the Nvidia GTX480 card, which is based on the Nvidia Fermi. 
We extended GPGPU-Sim to simulate a MT-CGRA core and a dMT-CGRA core and, using per-operation energy estimates obtained from RTL place\&route results for the new components, we extended the power model of GPUWattch to support the MT-CGRA and dMT-CGRA designs.

%%%%%%%%%%%%%%%%%%%%%%%%%%%%%%%%%%%%%%%%%%%%%%%%%%
\begin{table*}[t]
	%\vspace*{-1.5ex}
	\scriptsize
	\centering
	\begin{tabular}{| l | l | l | l |}
		\hline
		Application 	& Application Domain & Kernel	& Description  \\   \hline
		scan			& Data-Parallel Algorithms 	&\emph{scan\_naive}	& Prefix sum \\ %A strategy for searching in a graph \\
		matrixMul		& Linear Algebra       	&\emph{matrixMul}										& Matrix multiplication\\
		convolution   		& Linear Algebra  	&\emph{convolutionRowGPU}	& Convolution filter  \\%A solver for the three-dimensional Euler equations for compressible flow  \\
		reduce			& Data-Parallel Algorithms		&\emph{reduce} & Parallel Reduction\\
		%			&					&\emph{lud\_perimiter(22)}										& \\
		lud			& Linear Algebra		&\emph{lud\_internal} 	& Matrix decomposition\\% A solver for all of the variables in a linear system \\
		srad		& Ultrasonic/Radar Imaging &\emph{srad}									& Speckle Reducing Anisotropic Diffusion\\
		BPNN		& Pattern Recognition	& \emph{layerforward}				& Training of a neural network\\% A machine-learning algorithm that trains the weights of layered neural network 
		hotspot	& Physics Simulation  &\emph{hotspot\_kernel}										& Thermal simulation tool	\\
		pathfinder			& Dynamic Programming		&\emph{dynproc\_kernel}					& Find the shortest path on a 2-D grid\\
		\hline
	\end{tabular}
	\caption{A short description of the benchmarks that were used to evaluate the system\label{tab:rodinia_kernels}}
\end{table*}
%%%%%%%%%%%%%%%%%%%%%%%%%%%%%%%%%%%%%%%%%%%%%%%%%%

The system configuration is shown in Table~\ref{tab:cache_conf}.
By replacing the Fermi SM with a dMT-CGRA core, we retain the non-core components. The only difference between the processors' memory systems is that dMT-CGRA uses write-back and write-allocate policies in the L1 caches, as opposed to Fermi's write-through and write-no-allocate.

%\vspace{-1.5ex}
\paragraph*{\bf Compiler}\quad
We compiled CUDA kernels using LLVM~\cite{llvm} and extracted their SSA~\cite{cytron91} code. This was then used to configure the dMT-CGRA grid and interconnect. 

%\vspace{-1.5ex}
\paragraph*{\bf Benchmarks}\quad
We evaluated the dMT-CGRA architecture using kernels from the Nvidia SDK~\cite{nvidiasdk} and from the Rodinia benchmark suite~\cite{rodinia}, listed in Table~\ref{tab:rodinia_kernels}. We evaluated only kernels that use shared memory, and thus may benefit from our proposed programing model. 

%%%%%%%%%%%%%%%%%%%%%%%%%%%%%%%%%%%%%%%%%%%%%%%%%%%%%%%%%%%%%%%%%%%%%
\subsection{Simulation results}
%%%%%%%%%%%%%%%%%%%%%%%%%%%%%%%%%%%%%%%%%%%%%%%%%%%%%%%%%%%%%%%%%%%%%
This section evaluates the performance and power efficiency of the dMT-CGRA architecture. The results are compared to a baseline NVIDIA Fermi architecture and to an SGMF architecture, a multi threaded coarse grained reconfigurable array that does not support inter-thread communication (MT-CGRA). 

%%%%%%%%%%%%%%%%%%%%
%\vspace{-1ex}
\paragraph*{\bf Performance}\quad
%%%%%%%%%%%%%%%%%%%%

%%%%%%%%%%%%%%%%%%%%%%%%%%%%%%%%%%%%%%%%%%%%%%%%%%
\begin{figure}[t]
	\centering
	\vspace{-3ex}  
	\includegraphics[width=1\columnwidth]{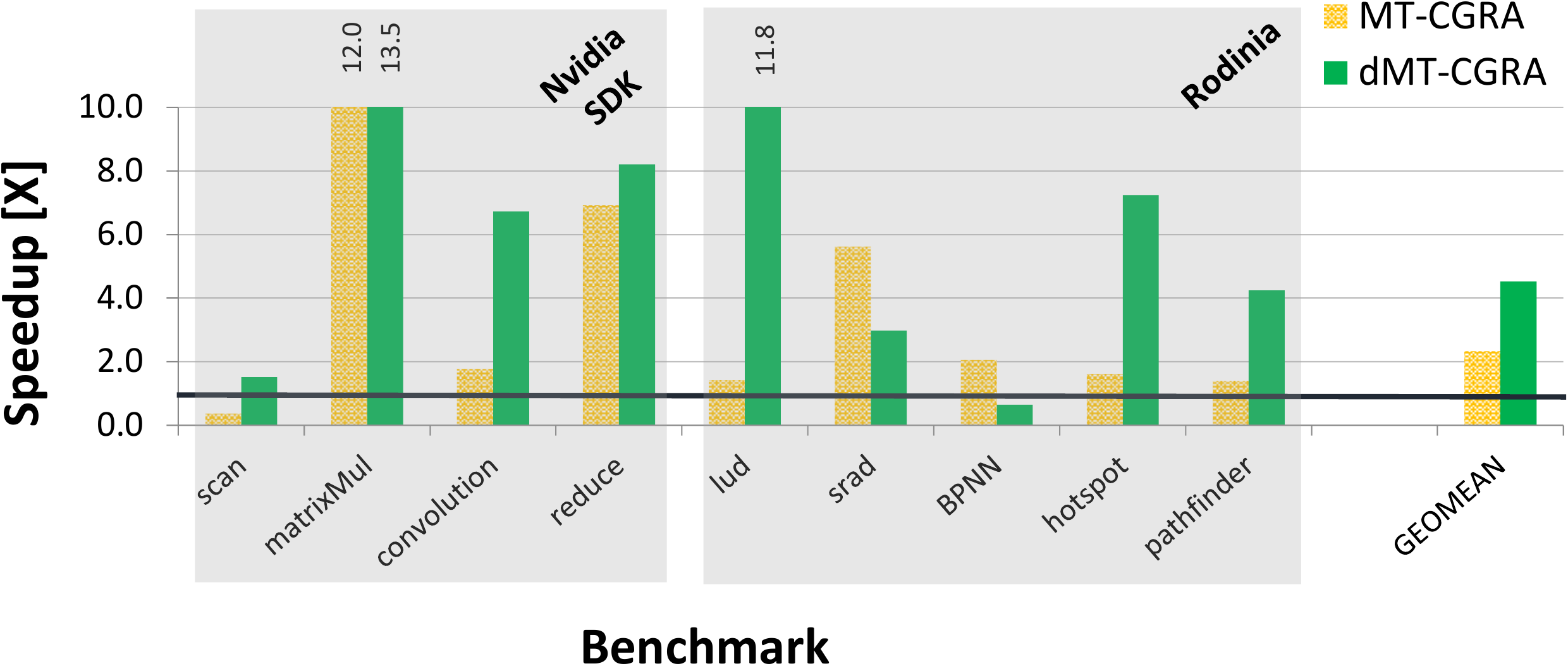}
	\caption{The speedup of the dMT-CGRA architecture and of a MT-CGRA architecture over the Fermi baseline  .\label{fig:speedup}}
\end{figure}
%%%%%%%%%%%%%%%%%%%%%%%%%%%%%%%%%%%%%%%%%%%%%%%%%% 

\fig{fig:speedup} demonstrates the performance improvements obtained from our proposed programing model and architecture over the Nvidia SDK and Rodinia kernels.

\fig{fig:speedup} demonstrates the speedup achieved by a dMT-CGRA core over a Fermi SM. While the performance improvement range varies across the different workloads, the combination of our programing model and adapted MT-CGRA architecture deliver speedups as high as 13.5$\times$\ with a geomean of 4.5$\times$. Spatial architectures, such as the dMT-CGRA and the MT-CGRA, are not bound by the width of the instruction fetch. Unlike GPUs, these architectures can operate all the functional units on the grid. Thus, a fully utilized spatial architecture composed of 140 units (such as the simulated dMT-CGRA and MT-CGRA) delivers a $\frac{140}{32}=4.375\times$ speedup over a fully utilize 32-wide GPU core. 

The MT-CGRA architecture suffers from a memory bottleneck and achieves an average speedup of 2.3$\times$, significantly lower than the speedup gained by the dMT-CGRA. The dMT-CGRA communicates intermediate values between threads and enables memory data reuse by different threads. Consequently,  the memory system BW is significantly reduced, usually eliminating the memory system bottleneck, thus enabling full ILP utilization.   

The results indicate that kernels originating from the Nvidia SDK gain significantly higher speedups than the kernels originating from the Rodinia benchmark suite. The kernels taken from the Nvidia SDK were re-implemented and their base algorithm was majorly revised to maximize the benefits from the new programing model and architecture proposed in this paper. Although the algorithms were altered, the functionally of those kernels was kept and the produced values are the same as of the original implementations. For the Rodinia kernels we used inter-thread communication instead of shared memory but tried to preserve the original algorithm as much as possible. The exceptions that stand out are the \emph{LUD} kernel in which we used our implementation of matrix multiplication; and \emph{scan}, a very sequential algorithm, in which inter-thread communication achieves significant energy reduction but without a significant speedup. Additionally, we preserved the original \emph{BPNN} kernel algorithm, resulting in an inter-thread communication slowdown of almost 40\%. The communication between adjacent threads limited the TLP and caused the slowdown. An implementation based on a different algorithm can potentially provide better performance. 

%%%%%%%%%%%%%%%%%%%%
%\vspace{-1ex}
\paragraph*{\bf Energy efficiency analysis}\quad
%%%%%%%%%%%%%%%%%%%%

%%%%%%%%%%%%%%%%%%%%%%%%%%%%%%%%%%%%%%%%%%%%%%%%%%
\begin{figure}[t]
\centering
\vspace*{-3ex}  
\includegraphics[width=1\columnwidth]{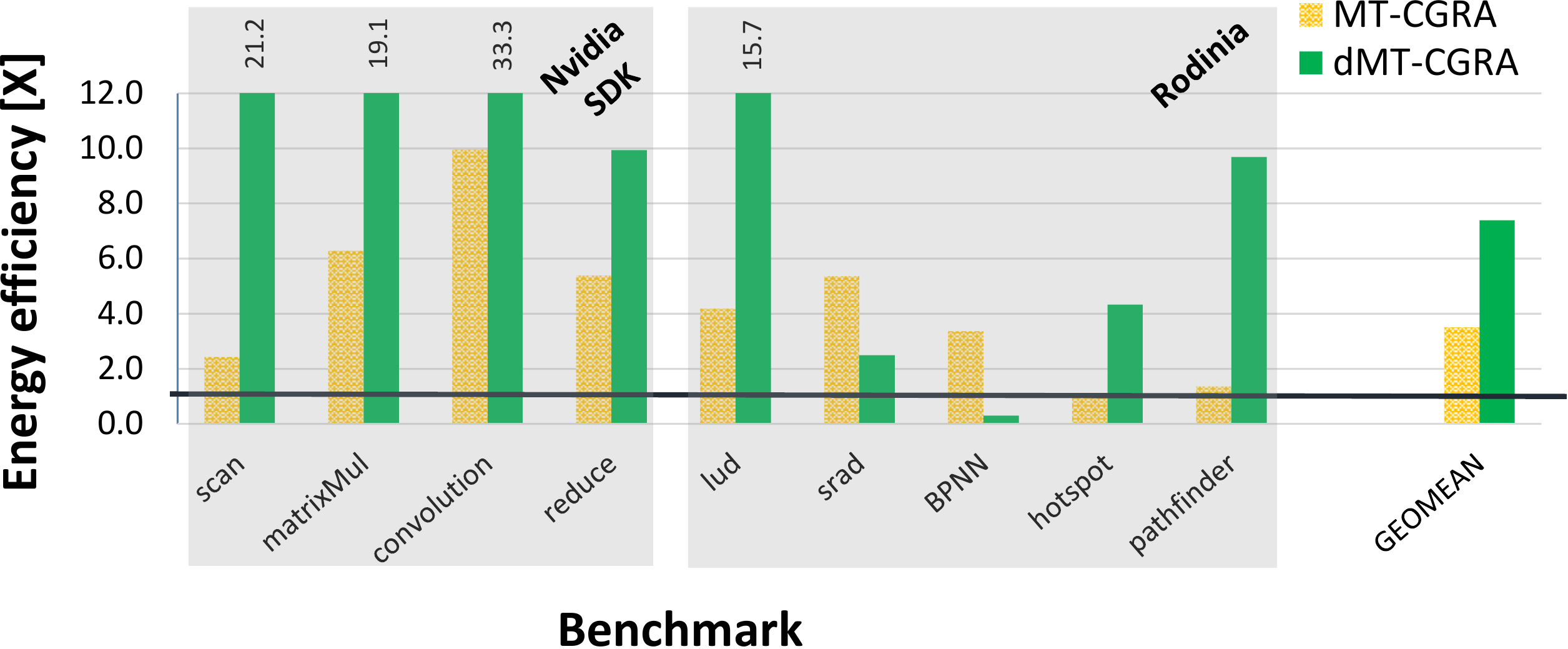}
\caption{Energy efficiency of a dMT-CGRA core over a MT-CGRA and Fermi SM.\label{fig:energy}}
\end{figure}
%%%%%%%%%%%%%%%%%%%%%%%%%%%%%%%%%%%%%%%%%%%%%%%%%% 

In this section we compare the energy efficiency of the evaluated architectures. Our evaluation compares the total energy required to complete the task, namely execute the kernel, since the different architectures use different \emph{instruction set architectures} (ISAs) and execute a different number of instructions for the same kernel. We simply multiply the execution time by the average power consumption for each architecture and divide it by the energy consumed by the Fermi baseline.

\fig{fig:energy} shows the energy efficiency of the dMT-CGRA and MT-CGRA architectures, compared to a Fermi GPU. As this figure demonstrates, the dMT-CGRA architecture is on average 7.4$\times$ more energy efficient while the MT-CGRA increases the energy efficiency by only 3.5$\times$. Our most outstanding case is the convolution kernel implementation using inter-thread communication. For this kernel the use of the dMT-CGRA programing model and architecture not only exposes available ILP and reduces the memory access overhead, it also significantly simplifies and shortens the code since no special treatment is needed for the margins, as demonstrated in \fig{fig:convolution}.  
Additionally, the Nvidia SDK implementations leverage the full benefits of the dMT-CGRA architecture and programing model and performed well. 

In the most cases, as expected, we see that high performance is usually translated into high energy efficiency, or in other words, tasks that terminate quickly consume less energy. The uncommon case is the  \emph{scan} kernel, where relative energy efficiency is extremely higher than its relative performance. That happens because its algorithm explicitly requires data transfers between threads. The programming model presented in this paper reduces the memory overhead for this kernel to the bare minimum.

To conclude, the evaluation demonstrates the performance and power benefits of the dMT-CGRA architecture and programing model over a von Neumann GPGPU (NVIDIA Fermi) and an MT-CGRA without the support of inter-thread communication.

	%%%%%%%%%%%%%%%%%%%%%%%%%%%%%%%%%%%%%%%%%%%%%%%%%%%%%%%%%%%%%%%%%%%%%%%%
\section{Related Work}
\label{sec:related}
%%%%%%%%%%%%%%%%%%%%%%%%%%%%%%%%%%%%%%%%%%%%%%%%%%%%%%%%%%%%%%%%%%%%%%%%

\paragraph*{\bf Dataflow architectures and CGRAs}\quad

There is a rich body of work regarding the potentials of dataflow based engines in general, and particularity coarse grained reconfigurable arrays. DySER~\cite{govindaraju11}, SEED~\cite{nowatzki15}, and MAD~\cite{ho15} extend von-Neumann based processors with dataflow engines that efficiently execute code blocks in a dataflow manner. Garp~\cite{callahan00} adds a CGRA component to a simple core in order to accelerate loops. While, TRIPS~\cite{sankaralingam03}, WaveScalar~\cite{swanson03} and Tartan~\cite{mishra06} portion the code into hyperblocks, schedule their execution according to the dependencies between them.   

However, these architectures mainly leverage their execution model to accelerate single threaded preference. With the exception of TRIPS~\cite{sankaralingam03} that enables multi threading by scheduling different threads to different tiles on the grid, and WaveCache~\cite{mishra06} that pipeline instances of hyperblocks originating from different threads. Nevertheless, neither of these architecture support simultaneous dynamic dataflow execution of threads on the same grid. While, SGMF~\cite{voitsechov14} and VGIW~\cite{voitsechov2015control} do support simultaneous dynamic multithreaded execution on the same grid, they do not support inter thread communication between threads.
\paragraph*{\bf Message passing and inter-core communication}\quad

Support of inter thread communication is vital when implementing efficient parallel software and algorithms. The MPI~\cite{mpi} programing model is perhaps the most scalable and commonly used message passing programing model. Many studies implemented hardware support for fine-grain communications across cores. The MIT Alewife machine~\cite{agarwal1995alewife}, MIT Raw~\cite{taylor02}, ADM~\cite{sanchez2010flexible}, CAF~\cite{wang2016caf} and the HELIX-RC architecture~\cite{campanoni2014helix} add an integrated hardware to multi-core systems, in order to provide fast communication and synchronization between the cores on chip. While,  XLOOPS~\cite{xloops} provides hardware mechanisms to transfer loop-carried dependencies across cores.  These prior works have explored hardware assisted techniques to add hardware support for communication between cores, in this paper we applied the same principles in a massively multithreaded environment and implemented communication between threads.

\paragraph*{\bf Inter-thread communication\bf }\quad

To enable decoupled software pipelining on sequential algorithms, DWSP~\cite{rangan2004decoupled} adds a synchronization buffer to support value communication between threads. Nvidia GPUs support inter thread communication within a warp using \emph{shuffle} instructions as described in the CUDA programing guide~\cite{cuda2015programming}, this form of communication is limited to data transfers within a warp and cannot be used in order to synchronize between threads since all thread within a warp execute in lockstep. Nevertheless, the addition of that instruction to the CUDA programing model, even in it's limited scope, demonstrates the need for inter thread communication in GPGPUs.

	%%%%%%%%%%%%%%%%%%%%%%%%%%%%%%%%%%%%%%%%%%%%%%%%%%%%%%%%%%%%%%%%%%%%%
\section{Conclusions}
\label{sec:conclusions}
%%%%%%%%%%%%%%%%%%%%%%%%%%%%%%%%%%%%%%%%%%%%%%%%%%%%%%%%%%%%%%%%%%%%%

Redundant memory access are a major bane for throughput processors. Such accesses can be  attributed to two major causes: using the memory (global or local) for inter-thread communication, and having multiple threads access the same memory locations.

In this paper we introduce direct inter-thread communication to the previously proposed \emph{multithreaded coarse-grain reconfigurable array} (MT-CGRA)~\cite{voitsechov14,voitsechov2015control}. The proposed \emph{dMT-CGRA} architecture eliminates redundant memory accesses by allowing threads to directly communicate through the CGRA fabric. The direct inter-thread communication
\begin{enumerate*}
  \item
    eliminates the use of memory as a communication medium; and
  \item
    allows thread to directly forward shared memory values rather than invoke redundant memory loads.
\end{enumerate*}

The elimination of redundant memory accesses both improve performance and reduce the energy consumption of the proposed dMT-CGRA architecture compared to MT-CGRA and NVIDIA GPGPUs. dMT-CGRA obtains average speedups of 1.95$\times$ and 4.5$\times$ over MT-CGRA and NVIDIA GPGPUs, respectively. At the same time, dMT-CGRA reduces energy consumption by an average of 53\% compared to MT-CGRA and 86\% compared to NVIDIA GPGPUs.

	%%%%%%% -- PAPER CONTENT ENDS -- %%%%%%%%
	
	%%%%%%% -- PAPER CONTENT ENDS -- %%%%%%%%
	
	%%%%%%%%% -- BIB STYLE AND FILE -- %%%%%%%%
	\bibliographystyle{ieeetr}
	\bibliography{ref}

\begin{thebibliography}{10}

\bibitem{googleTPU}
N.~P. Jouppi, C.~Young, N.~Patil, D.~Patterson, G.~Agrawal, R.~Bajwa, S.~Bates,
  S.~Bhatia, N.~Boden, A.~Borchers, R.~Boyle, P.-l. Cantin, C.~Chao, C.~Clark,
  J.~Coriell, M.~Daley, M.~Dau, J.~Dean, B.~Gelb, T.~V. Ghaemmaghami,
  R.~Gottipati, W.~Gulland, R.~Hagmann, C.~R. Ho, D.~Hogberg, J.~Hu, R.~Hundt,
  D.~Hurt, J.~Ibarz, A.~Jaffey, A.~Jaworski, A.~Kaplan, H.~Khaitan,
  D.~Killebrew, A.~Koch, N.~Kumar, S.~Lacy, J.~Laudon, J.~Law, D.~Le, C.~Leary,
  Z.~Liu, K.~Lucke, A.~Lundin, G.~MacKean, A.~Maggiore, M.~Mahony, K.~Miller,
  R.~Nagarajan, R.~Narayanaswami, R.~Ni, K.~Nix, T.~Norrie, M.~Omernick,
  N.~Penukonda, A.~Phelps, J.~Ross, M.~Ross, A.~Salek, E.~Samadiani, C.~Severn,
  G.~Sizikov, M.~Snelham, J.~Souter, D.~Steinberg, A.~Swing, M.~Tan,
  G.~Thorson, B.~Tian, H.~Toma, E.~Tuttle, V.~Vasudevan, R.~Walter, W.~Wang,
  E.~Wilcox, and D.~H. Yoon, ``In-datacenter performance analysis of a tensor
  processing unit,'' in {\em Intl.\ Symp.\ on Computer Architecture (ISCA)},
  2017.

\bibitem{wavecomputingWP}
C.~Nicol, ``A coarse grain reconfigurable array (cgra) for statically scheduled
  data flow computing,'' tech. rep., Wave Computing, 2017.

\bibitem{govindaraju11}
V.~Govindaraju, C.-H. Ho, and K.~Sankaralingam, ``Dynamically specialized
  datapaths for energy efficient computing,'' in {\em Symp.\ on
  High-Performance Computer Architecture (HPCA)}, Feb 2011.

\bibitem{chen2017eyeriss}
Y.-H. Chen, T.~Krishna, J.~S. Emer, and V.~Sze, ``Eyeriss: An energy-efficient
  reconfigurable accelerator for deep convolutional neural networks,'' {\em
  IEEE Journal of Solid-State Circuits}, vol.~52, no.~1, 2017.

\bibitem{nowatzki2016pushing}
T.~Nowatzki, V.~Gangadhan, K.~Sankaralingam, and G.~Wright, ``Pushing the
  limits of accelerator efficiency while retaining programmability,'' in {\em
  Symp.\ on High-Performance Computer Architecture (HPCA)}, 2016.

\bibitem{voitsechov14}
D.~Voitsechov and Y.~Etsion, ``Single-graph multiple flows: Energy efficient
  design alternative for {GPGPUs},'' in {\em Intl.\ Symp.\ on Computer
  Architecture (ISCA)}, 2014.

\bibitem{voitsechov2015control}
D.~Voitsechov and Y.~Etsion, ``Control flow coalescing on a hybrid dataflow/von
  {Neumann} {GPGPU},'' in {\em Intl.\ Symp.\ on Microarchitecture (MICRO)},
  2015.

\bibitem{podlozhnyuk2007image}
V.~Podlozhnyuk, ``Image convolution with {CUDA},'' tech. rep., NVIDIA, Jun
  2007.

\bibitem{nvidiasdk}
{NVIDIA}, ``{CUDA} {SDK} code samples.''

\bibitem{cuda2015programming}
NVIDIA, {\em CUDA Programming Guide v7.0}, Mar 2015.

\bibitem{arvind90}
Arvind and R.~Nikhil, ``Executing a program on the {MIT} tagged-token dataflow
  architecture,'' {\em IEEE Trans.\ on Computers}, vol.~39, pp.~300--318, Mar
  1990.

\bibitem{patt85}
Y.~N. Patt, W.~M. Hwu, and M.~Shebanow, ``{HPS}, a new microarchitecture:
  rationale and introduction,'' in {\em Intl.\ Symp.\ on Microarchitecture
  (MICRO)}, 1985.

\bibitem{gpgpusim}
A.~Bakhoda, G.~L. Yuan, W.~W.~L. Fung, H.~Wong, and T.~M. Aamodt, ``Analyzing
  {CUDA} workloads using a detailed {GPU} simulator.,'' in {\em IEEE Intl.\
  Symp.\ on Perf.\ Analysis of Systems and Software (ISPASS)}, 2009.

\bibitem{leng13}
J.~Leng, T.~Hetherington, A.~ElTantawy, S.~Gilani, N.~S. Kim, T.~M. Aamodt, and
  V.~J. Reddi, ``{GPUWattch}: enabling energy optimizations in {GPGPUs},'' in
  {\em Intl.\ Symp.\ on Computer Architecture (ISCA)}, 2013.

\bibitem{llvm}
C.~Lattner and V.~Adve, ``{LLVM}: A compilation framework for lifelong program
  analysis \& transformation,'' in {\em Intl.\ Symp.\ on Code Generation and
  Optimization (CGO)}, 2004.

\bibitem{cytron91}
R.~Cytron, J.~Ferrante, B.~K. Rosen, M.~N. Wegman, and F.~K. Zadeck,
  ``Efficiently computing static single assignment form and the control
  dependence graph,'' {\em ACM Trans.\ on Programming Languages and Systems},
  vol.~13, no.~4, pp.~451--490, 1991.

\bibitem{rodinia}
S.~Che, M.~Boyer, J.~Meng, D.~Tarjan, J.~W. Sheaffer, S.-H. Lee, and
  K.~Skadron, ``Rodinia: A benchmark suite for heterogeneous computing,'' in
  {\em IEEE Intl.\ Symp.\ on Workload Characterization (IISWC)}, 2009.

\bibitem{nowatzki15}
T.~Nowatzki, V.~Gangadhar, and K.~Sankaralingam, ``Exploring the potential of
  heterogeneous von neumann/dataflow execution models,'' in {\em Intl.\ Symp.\
  on Computer Architecture (ISCA)}, ACM, Jun 2015.

\bibitem{ho15}
C.-H. Ho, S.~J. Kim, and K.~Sankaralingam, ``Efficient execution of memory
  access phases using dataflow specialization,'' in {\em Intl.\ Symp.\ on
  Computer Architecture (ISCA)}, Jun 2015.

\bibitem{callahan00}
T.~J. Callahan and J.~Wawrzynek, ``Adapting software pipelining for
  reconfigurable computing,'' in {\em Intl.\ Conf.\ on Compilers, Architecture,
  and Synthesis for Embedded Systems}, 2000.

\bibitem{sankaralingam03}
K.~Sankaralingam, R.~Nagarajan, H.~Liu, C.~Kim, J.~Huh, D.~Burger, S.~W.
  Keckler, and C.~R. Moore, ``Exploiting {ILP}, {TLP}, and {DLP} with the
  polymorphous {TRIPS} architecture,'' in {\em Intl.\ Symp.\ on Computer
  Architecture (ISCA)}, 2003.

\bibitem{swanson03}
S.~Swanson, K.~Michelson, A.~Schwerin, and M.~Oskin, ``{WaveScalar},'' in {\em
  Intl.\ Symp.\ on Microarchitecture (MICRO)}, Dec 2003.

\bibitem{mishra06}
M.~Mishra, T.~J. Callahan, T.~Chelcea, G.~Venkataramani, M.~Budiu, and S.~C.
  Goldstein, ``{Tartan}: Evaluating spatial computation for whole program
  execution,'' in {\em Intl.\ Conf.\ on Arch.\ Support for Prog.\ Lang.\ \&
  Operating Systems (ASPLOS)}, Oct 2006.

\bibitem{mpi}
{Message Passing Interface Forum}, ``{MPI}: A message-passing interface
  standard,'' Jun 2015.
\newblock Version 3.1.

\bibitem{agarwal1995alewife}
A.~Agarwal, R.~Bianchini, D.~Chaiken, K.~L. Johnson, D.~Kranz, J.~Kubiatowicz,
  B.-H. Lim, K.~Mackenzie, and D.~Yeung, ``The {MIT} alewife machine:
  Architecture and performance,'' in {\em Intl.\ Symp.\ on Computer
  Architecture (ISCA)}, 1995.

\bibitem{taylor02}
M.~Taylor, J.~Kim, J.~Miller, D.~Wentzlaff, F.~Ghodrat, B.~Greenwald,
  H.~Hoffman, P.~Johnson, J.-W. Lee, W.~Lee, A.~Ma, A.~Saraf, M.~Seneski,
  N.~Shnidman, V.~Strumpen, M.~Frank, S.~Amarasinghe, and A.~Agarwal, ``The
  {Raw} microprocessor: a computational fabric for software circuits and
  general-purpose programs,'' {\em IEEE Micro}, vol.~22, no.~2, 2002.

\bibitem{sanchez2010flexible}
D.~Sanchez, R.~M. Yoo, and C.~Kozyrakis, ``Flexible architectural support for
  fine-grain scheduling,'' in {\em Intl.\ Conf.\ on Arch.\ Support for Prog.\
  Lang.\ \& Operating Systems (ASPLOS)}, 2010.

\bibitem{wang2016caf}
Y.~Wang, R.~Wang, A.~Herdrich, J.~Tsai, and Y.~Solihin, ``{CAF}: Core to core
  communication acceleration framework,'' in {\em Intl.\ Conf.\ on Parallel
  Arch.\ and Compilation Techniques (PACT)}, 2016.

\bibitem{campanoni2014helix}
S.~Campanoni, K.~Brownell, S.~Kanev, T.~M. Jones, G.-Y. Wei, and D.~Brooks,
  ``{HELIX-RC}: An architecture-compiler co-design for automatic
  parallelization of irregular programs,'' in {\em Intl.\ Symp.\ on Computer
  Architecture (ISCA)}, 2014.

\bibitem{xloops}
S.~Srinath, B.~Ilbeyi, M.~Tan, G.~Liu, Z.~Zhang, and C.~Batten, ``Architectural
  specialization for inter-iteration loop dependence patterns,'' in {\em Intl.\
  Symp.\ on Microarchitecture (MICRO)}, Dec 2014.

\bibitem{rangan2004decoupled}
R.~Rangan, N.~Vachharajani, M.~Vachharajani, and D.~I. August, ``Decoupled
  software pipelining with the synchronization array,'' in {\em Intl.\ Conf.\
  on Parallel Arch.\ and Compilation Techniques (PACT)}, 2004.

\end{thebibliography}
	%%%%%%%%%%%%%%%%%%%%%%%%%%%%%%%%%%%%
	
\end{document}